\documentclass[%
 reprint,
 amsmath,amssymb,
 aps,
prb,
]{revtex4-2}
\DeclareMathOperator{\sgn}{sign}

\usepackage{graphicx}
\usepackage{dcolumn}
\usepackage{bm}
\usepackage{hyperref}

\begin{document}

\preprint{APS/123-QED}

\title{Finite and High-temperature series expansion via many-body perturbation theory}

\author{Mohamed Amine Tag}
\email{mohamed.tag@univ-tebessa.dz}
\author{Abid Boudiar}%
\author{Mohamed El-Hadi Mansour}%
\author{Abdelkader Hafdallah}%
\author{Chafia Bendjeroudib}%
\affiliation{%
 Applied and Theoretical Physics Laboratory (LPAT) \\
 Larbi Tebessi University- Tebessa, Algeria  
}%




\date{\today}

\begin{abstract}
We present a new algorithm to evaluate the grand potential at finite and high-temperature series expansion via many-body perturbation theory. This algorithm allows us to formulate each order as a divided difference. Further, we apply this algorithm to the Heisenberg spin-1/2 XXZ chain. We obtain all coefficients of the high-temperature expansion of the free energy and susceptibility per site of this model up to sixth order.
\end{abstract}

\maketitle


\section{\label{sec:level1} Introduction}
The study of many-body Hamiltonian for strongly interacting particles is the most complicated problem in condensed matter physics due to the absence of exact solutions except in some cases like a one-dimensional interacting model. For this, the Hamiltonian is treated in another way. The most important method is the many-body perturbation theory (MBPT) in finite temperature, which is an infinite sum of orders in terms of interacting potential. The contribution of each term is simplified thanks to the Feynman diagrammatic formalism. Still, this method poses some difficulties due to the factorial growth of the diagrams number \cite{Kugler2018} and to the numerical integration of each term.
\\One of the most used methods for numerically evaluating the contribution of each Feynman diagram is diagrammatic Monte Carlo (DiagMC) \cite{Prokof1998,Prokof2008,Houcke2010,Kozik2010,Houcke2012,Rossi2018}. This method has been developed, and one of the powerful simplifications is to deal with the Hugenholtz diagrams \cite{Hug1965}, which is the combination between direct and exchange potential, instead of ordinary Feynman diagrams \cite{Chen2019,bao2021}. This improvement has two advantages. First, it is directly processed with the essentially distinct diagram (EDD) chosen from its equivalent diagrams because the standard DiagMC method samples the diagrams one by one. Second, these smaller numbers of Hugenholtz diagrams increase the efficiency of Monte Carlo sampling and decrease the computation cost. In this case, our previous fast algorithm \cite{tag2017} could be used to generate all EDDs.    
\\The conventional DiagMC relies on the Matsubara formalism, but recently, a new algorithm was proposed to simplify the evaluation of diagrams \cite{tag2019,taheridehkordi2019,taheridehkordi2020}. In particular, this development revolves around reducing the integration of diagrams in space-time to only space. Our method \cite{tag2019} uses some basic definitions in graph theory, whereas the work of \cite{taheridehkordi2019,taheridehkordi2020} is a direct application of residue theorem to each diagram. These methods reduce the calculation time and allow us to evaluate the grand potential or self-energy in the finite temperature of each order in MBPT.
\\High-temperature series expansions (HTS) is a traditional method that can be used to treat any problem in thermodynamic. The previously powerful methods like finite cluster method \cite{Oitmaa1996,Gelfand1990,Gelfand2000}, linked-cluster expansions \cite{Oitmaa2006,Sykes1966}, finite lattice method \cite{Enting1996,Neef1977} and the numerical linked-cluster expansions \cite{Rigol2006,Khatami2011} are based on the Taylor expansion of the partition function around $\beta (\beta=1/T)$. However, all these methods have a limited series, and the progress is mainly due to improving the performance of computers.
\\In this manuscript, we propose a new general method that allows us to evaluate the contribution of each vacuum diagram so that the result is adequate with the finite and high-temperature series expansion. For example, we calculate the free energy of the Heisenberg spin ½ XXZ chain up to the sixth order and compare the results to other methods.




\section{The model}
The general expression of the interacting Hamiltonian in the second quantification is expressed as follows.
\begin{equation}
H=H_{0}+H_{I}=\sum_{p}E_{p}a_{p}^{+}a_{p}+\frac{1}{4}\sum_{rsml}V_{m,l}^{r,s}a_{r}^{+}a_{s}^{+}a_{l}a_{m}
\label{eq:1}.
\end{equation}

\begin{figure}[b]
	\includegraphics[scale = 0.65]{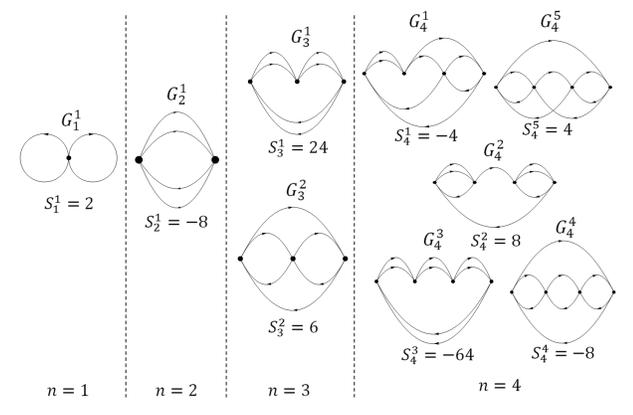}
	\caption{\label{fig:img0} Example of connected vacuum diagrams $G_{n}^{i}$ with its symmetry factors of order 1,2, 3 and 4.}
\end{figure}

Where the energy $E_{k}$  is the combination of quasi-particle energy $\varepsilon_{k}$ and the chemical potential $\mu$:
\begin{equation}
    E_{k}=\varepsilon_{k}-\mu
    \label{eq:2}.
\end{equation}
The interacting potential $V_{m,l}^{r,s}$ represented as the sum of direct $\langle rs|V|ml\rangle$ and exchange  potentials $\langle rs|V|lm\rangle$.
\begin{equation}
    V_{m,l}^{r,s}=\langle rs|V|ml\rangle+\epsilon \langle rs|V|lm\rangle
    \label{eq:3}.
\end{equation}
The constant $\epsilon$ in Eq.~(\ref{eq:3}) equal to -1 for fermions and +1 for bosons.
In the Hamiltonian Eq.~(\ref{eq:1}), $a_{p}^{+}$ and $a_{p}$ are respectively the creation and annihilation operators.
The series expansion of the grand potential $\Omega$ in MBPT can be expressed as the sum of order $n$ of $\Omega_{n}$
\begin{equation}
    \Omega=\Omega_{0}+\sum_{n=1}^{\infty}\Omega_{n}
    \label{eq:4}.
\end{equation}
Each $\Omega_{n}$ contains the contribution of all Essentially Distinct Hugenholtz Diagrams (EDDs) at order $n$ (EDD is the diagram chosen among the equivalent diagrams, for more information, see Ref.~[\onlinecite{tag2017}]). For simplicity, we select only the graphs that don’t contain the Hartree-Fock circle, this means that any line of diagrams already includes the sub Hartree-Fock contribution, and further, we replace the energy by:
\begin{equation}
    E_{p}=\varepsilon_{p}-\mu-e_{p}^{HF}(\beta)
    \label{eq:5}.
\end{equation}
Where $e_{p}^{HF}(\beta)$ is the Hartree-Fock energy which depends on the inverse of temperature $\beta$. We can find the expression of this energy by solving the following non-linear integral equation:
\begin{equation}
    e_{p}^{HF}(\beta)=\int V_{p,q}^{p,q}f^{-}(E_{q})dq
    \label{eq:6}.
\end{equation}
Where $f^{-}$ represent the statistical factor:
\begin{equation}
    f^{-}(E_{p})=\frac{\epsilon}{e^{\beta E_{p}}-\epsilon}
    \label{eq:7}.
\end{equation}
The contribution of $\Omega_{n}$ formulated as:
\begin{equation}
    \Omega_{n}=\sum_{i=1}^{N}\frac{\Omega_{n}^{G_i}}{S_{i}}
    \label{eq:8}.
\end{equation}
Where $\Omega_{n}^{G_{i}}$ represents the contribution of diagram $G_{i}$ to the grand potential and $S_{i}$ is the symmetry factor of this diagram. Fig.~(\ref{fig:img0}) shows an example of conventional vacuum diagrams $G_{i}$.
\\The number $N$ of vacuum Hugenholtz diagrams $G_{i}$ up to order 10 shown in Table~\ref{tab:tb1}.
\begin{table}[b]
\caption{\label{tab:tb1}%
The total number of distinct vacuum Hugenholtz-type diagrams without Hartree-Fock sub-diagrams.
}
\begin{ruledtabular}
\begin{tabular}{lcdr}
\textrm{n}&
\textrm{N}\\
\colrule
1 & 1\\
2 & 1\\
3 & 2\\
4 & 5\\
5 & 13\\
6 & 59\\
7 & 285\\
8 & 1987\\
9 & 16057\\
10 & 149430\\
\end{tabular}
\end{ruledtabular}
\end{table}
Each $\Omega_{n}^{G_{i}}$ consists of summations over momenta within the first Brillouin zone and Matsubara frequencies. In previous work Ref.~[\onlinecite{tag2019}], we developed an algorithm to evaluate the Matsubara sums based on some basic definitions in graph theory. This method allows us to formulate the contribution $\Omega_{n}^{G_{i}}$ of each graph $G_{i}$ in terms of statistical factors $f^{-}$, $f^{+}=1+f^{-}$ and energies. We call this method the graphical evaluation of Matsubara sums.

\subsection{Graphical evaluation of Matsubara sums}
\begin{figure}[b]
\includegraphics[scale = 0.55]{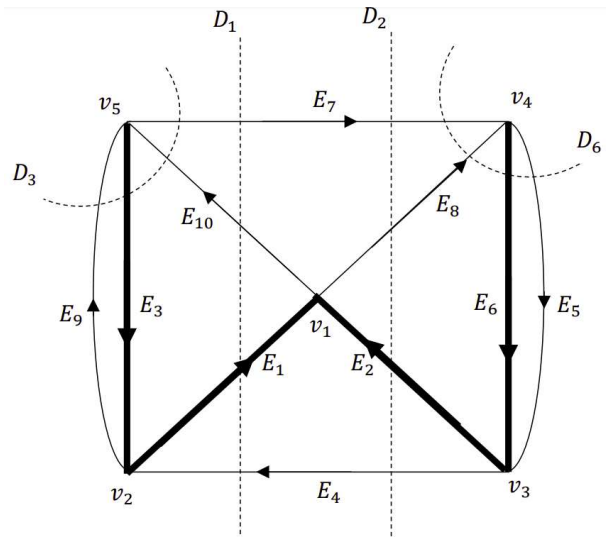}
\caption{\label{fig:figu1} A connected graph G(5,10) with an example of a spanning tree $T$ (Thick lines), and the
associated co-tree $T^*$ (thin lines). The dashed lines represent the fundamental cuts.}
\end{figure}

This method helps us calculate the summation over Matsubara frequencies of any distinct Hugenholtz (or Feynman) diagrams. 
The distinct Hugenholtz diagram is a connected graph $G(n,2n)$ with $n$ vertices and $2n$ edges selected from its equivalent. 
\\Before treating this method, we recall some basic definitions in graph theory:
\begin{enumerate}
\item The spanning tree $T$ is a tree that includes all the vertices of $G$ without any cycles. The $n-1$ edges of this spanning tree are called branches.
\item The complement of a spanning tree called co-tree $T^{*}$. All $n+1$ edges of co-tree are called chords.
\item Each edge (line) of the connected graph is associated with a coefficient, which is an integer number, $n_{i}>0,1\leq i \leq 2n$.
\end{enumerate}
The graph $G$ contains a large number of spanning trees (in the worst case, the number of all spanning trees in Hugenholtz diagram $G(n,2n)$ is $4^{n-1}$).
\\In this method, the general formula of the contribution $\Omega_{n}^{G_{i}}$ of a given connected diagram $G_{i}$ to the grand potential can be formulated as:
\begin{equation}
    \Omega_{n}^{G_{i}}=\sum_{T}\frac{N_{T^{*}}}{D_{T}}
    \label{eq:9}.
\end{equation}
The sum $\sum_{T}$ in Eq.~(\ref{eq:9}) is over all spanning trees. For each spanning tree $T$,the denominator $D_{T}$ defined as:
\begin{equation}
    D_{T}=\prod_{i=1}^{n-1}D_{i}
    \label{eq:10}.
\end{equation}
Where $D_i$ is the contribution of each branch $i$ of the spanning tree $T$  
\begin{equation}
    D_i=E_i+\sum E_{c_1}-\sum E_{c_2}
    \label{eq:11}.
\end{equation}
Where $E_{i}$ is the energy of the branch $i$.
$\sum E_{c_1}$($\sum E_{c_2}$) is the sum of the energies of the chords which are in the same direction (opposite direction) as the energy of the branch $E_i$. Here $E_i \in T$ and $E_{c_1},E_{c_2}\in T^{*}$. In graph theory, the set of $D_i$ are called the fundamental cuts.\\
In \ref{eq:9}, the numerator $N_{T^{*}}$ represent the contribution of the co-tree $T^{*}$ and defined by the following relation:
\begin{equation}
    N_{T^{*}}=\prod_{j=1}^{n+1}f^{[O_j]}(E_j)
    \label{eq:12}.
\end{equation}
Where $f^{[O_j]}(E_j)$ is the statistical factor of each chord $j$ of co-tree $T^*$. Here $E_j \in T^*$.
\\The amount $[O_j] =\sgn (O_j) = \pm$ is the sign of the total orientation $O_j$ of each chord $j$. The latter can be formulated as:
\begin{equation}
    O_j=n_j+\sum n_{b_1}-\sum n_{b_2}
    \label{eq:13}.
\end{equation}
Where $n_j > 0$ is the coefficient of the chord $j$.
\\The sum $\sum n_{b_1}$ is the total number of coefficients of the branches encountered in the same direction as $n_j$, whereas $\sum n_{b_1}$ is the total number of coefficients of the branches pointing in the opposite direction to $n_j$. Here $n_j \in T^*$ and $n_{b_1},n_{b_2} \in T$. In graph theory, the set of $O_j$ called the fundamental cycles.
\\Finally, the integer coefficients $ n_j> 0 $ are chosen arbitrarily so that the total orientations $O_j$ of all the cycles of the graph $G$ are not equal to zero.
\\Fig.~(\ref{fig:figu1}) shows an example of connected Hugenholtz diagram $G$ with vertices $v_i;1\leq i \leq 5$ and edges $E_i;1\leq i\leq 10$. The arbitrary spanning tree $T$ defined by the branches $E_1,E_2,E_3$ and $E_6$ (thick lines) and the associated co-tree $T^*$ represented by the chords $E_4,E_5,E_7,E_8,E_9$ and $E_{10}$ (thin lines). The contribution of this spanning tree is the fraction $N_{T^*}/D_T$, where the denominator $D_T$ can be evaluated from the fundamental cutsets $D=D_1D_2D_3D_6$ and the numerator $N_{T^*}$ determined from the chords of co-tree $N_{T^*}=f_{4}^{[O_4]}f_{5}^{[O_5]}f_{7}^{[O_7]}f_{8}^{[O_8]}f_{9}^{[O_9]}f_{10}^{[O_{10}]}$. All cutsets are simply extracted from the graph: $D_1=E_1+E_7-E_4-E_{10}$, $D_2=E_2+E_4-E_7-E_8$, $D_3=E_3+E_7-E_9-E_{10}$ and $D_6=E_6+E_5-E_7-E_8$. The total orientations $O_i$ can be calculated directly from the fundamental cycles of the graph which are $O_4=n_4+n_1-n_2$, $O_5=n_5-n_6$, $O_7=n_7+n_6+n_2-n_1-n_3$, $O_8=n_8+n_6+n_2$, $O_9=n_9+n_3$ and $O_{10}=n_{10}+n_3+n_1$.

The arbitrary integer values of the edge coefficients $n_i > 0$ can be determined by finding all the cycles of the directed graph $G$.
\\For more details on the graphical evaluation of Matsubara sums and its algorithms, see our previous work Ref.~[\onlinecite{tag2019}].

\section{Formulate the expansion of grand potential as a divided difference}

This part will discuss how to reformulate the expansion  Eq.~(\ref{eq:9}) in the form of divided difference. We will show how this new reformulation helps us find the high temperatures series expansion of the grand potential efficiently. We will limit the study to fermions (the constant $\epsilon=-1$ in Eq.~(\ref{eq:3})) (of course they can be generalized to bosons).
\\Before continuing to the method, let us begin with the second order's simplest case.
\\The contribution of diagram $G_{2}^{1}$ (Fig.~(\ref{fig:img0})) to free energy is given by the formula: 
\begin{eqnarray}
\begin{aligned}
    \Omega_2&=-\frac{(f^-(x_2)(f^-(x_3)+f^+(x_4))-f^-(x_3)f^-(x_4))f^-(x_1)}{x_1+x_2-x_3-x_4}\\
     &+\frac{f^+(x_2)f^-(x_3)f^-(x_4)}{x_1+x_2-x_3-x_4}
     \end{aligned}
     \label{eq:14}.
\end{eqnarray}
Where $x_i = E_{p_i}$. Replacing the nominator of second therm in RHS of relation Eq.~(\ref{eq:14}) by the following identity:
\begin{eqnarray}
\begin{aligned}
&\frac{f^+(x_2)f^-(x_3)f^-(x_4)}{f^-(-x_2+x_3+x_4)}=\\
&f^-(x_2)(f^-(x_3)+f^+(x_4))-f^-(x_3)f^-(x_4)
\label{eq:15}.
\end{aligned}
\end{eqnarray}
Then the form of $\Omega _2$ can be rewritten as
\begin{eqnarray}
\begin{aligned}
    &\Omega_2=-(f^-(x_2)(f^-(x_3)+f^+(x_4))-f^-(x_3)f^-(x_4))\\
    &\times f^-[x_1,-x_2+x_3+x_4]
    \label{eq:16}.
\end{aligned}
\end{eqnarray}
Where the amount $f^-[x_1,-x_2+x_3+x_4]$ is the first order divided difference and defined by the fraction:
\begin{equation}
    f^-[x_1,-x_2+x_3+x_4]=\frac{f^-(x_1)-f^-(-x_2+x_3+x_4)}{x_1+x_2-x_3-x_4}
    \label{eq:17}.
\end{equation}
The relation Eq.~(\ref{eq:16}) helps us to find the HTS of $\Omega _2$ because the ratio Eq.~(\ref{eq:17}) can easily be written as an infinite sum in terms of the inverse temperature $\beta$, i.e.,
\begin{equation}
    A^{(1)}(x,y)=-\frac{1}{2}\sum_{n=1}^{\infty}\frac{\beta^n}{n!}e_n\sum_{k=0}^{n-1}x^{k}y^{n-1-k}
    \label{eq:18}.
\end{equation}
Where the coefficients $e_n$ are the Euler constants
\begin{equation}
    e_n=\frac{\partial ^n}{\partial t^n}\left(\frac{2}{1+e^t}\right)_{t=0};n\geq 1
    \label{eq:19}.
\end{equation}
The previous writing of $\Omega_n$ in Eq.~(\ref{eq:16}) can be deduced by performing the Euclidean division of the fraction $f^-(x_1)/ (x_1 + x_2-x_3-x_4)$ of the first part in RHS of Eq.~(\ref{eq:14}), in which the variable $x_1==E_{p_1}$ is the divisor. In this division operation, we only need the quotient while the remainder will vanish automatically with the second part in RHS of Eq.~(\ref{eq:14}).
\\Before starting to process the higher order of $\Omega _n$, we note that in the relation Eq.~(\ref{eq:9}), the fraction numerator contains the statistical factors $f^-$ or $f^+$ which they are in terms of $\beta$. When expanding these statistical coefficients at high temperatures $\beta <<1$ to desired order $R$, the numerator $N_{T^*}$ will be a polynomial of degree $R$ in terms of $\beta$ as well the energies $E_j$. Therefore, the Taylor expansion of $N_{T^*}$ can be written as a multivariate polynomial of energies
\begin{equation}
    N_{T^*}=\sum_{k=0}^{R} \left(\frac{-\beta}{2} \right)^k \prod_{j=1}^{n+1}\frac{E_{j}^{k_j}}{k_{j}!}e_{k_j} 
    \label{eq:20}.
\end{equation}
Where the addition coefficients $k_j$, which are positive integers, are chosen by solving the integer equation $\sum_{j=1}^{n+1}k_j=k$.
\\In the case $k_j=0$, the coefficient $e_0(0)$ equal to the sign of total orientation $e_0=-[O_j]$.
\\As shown by the fraction of relation Eq.~(\ref{eq:9}), the numerator Eq.~(\ref{eq:20}) must be within the limits of denominator Eq.~(\ref{eq:10}) in terms of the multivariate polynomial degree of energies. In other words, the degree of $N_{T^*}$, must be greater than $n-2$, while the sum over all spanning trees of the coefficients $k$ which is less than $n-1$, must be equal to zero:
\begin{equation}
    \sum_{T}\frac{\prod_{j=1}^{n+1}\frac{E_{j}^{k_j}}{k_{j}!}e_{k_j}}{\prod_{i=1}^{n-1}D_i}=0;0\leq k\le n-2
    \label{eq:21}.
\end{equation}
In resuming, from the relation Eq.~(\ref{eq:20}) and Eq.~(\ref{eq:21}),the grand potential Eq.~(\ref{eq:9}) at high temperature can be written as the following sum of fraction
\begin{equation}
    \Omega_{n}^{G_i}=\sum_{k=n-1}^{R}\beta^k\sum_{T}\frac{P_k(E_1,...,E_{n+1})}{P_{n-1}(E_1,...,E_{n-1})}
    \label{eq:22}.
\end{equation}
Where the multivariate polynomials $P_k(E_1,...,E_{n+1})$ and $P_{n-1}(E_1,...,E_{n-1})$ are written in the following forms:
\begin{equation}
    P_k(E_1,...,E_{n+1})=\prod_{j=1}^{n+1}\frac{E_{j}^{k_j}}{k_{j}!}e_{k_j}
    \label{eq:23}.
\end{equation}
\begin{equation}
   P_{n-1}(E_1,...,E_{n-1})=\prod_{i=1}^{n-1}D_i
   \label{eq:24}.
\end{equation}
The direct application of relation Eq.~(\ref{eq:22}) takes more time because we simplify all the fractions. This process becomes more complicated in the cases of diagrams that contain many spanning trees.
\\Since the numerator and denominator of Eq.~(\ref{eq:22}) are multivariate polynomials in terms of energies; therefore, we do the euclidean division of polynomials. Additionally, the final expression of the grand potential $\Omega_{n}^{G_i}$ will be a multivariate polynomial in terms of energies and cannot contain any fraction. In this case, we only need the quotient of the division process (the sum over spanning trees $T$ of the remainders will be zero). In this way, we take an arbitrary variable of energy, for example $E_r$, and then we perform the polynomials division $P_k(E_r)/P_{n-1}(E_r)$ on all spanning trees $T$. If the quotient of division still contains a fraction of another polynomial, say $P_k(E_s)/P_{n-1}(E_s)$, we repeat the division operation for $E_s$, and so on until we get the expression of grand potential in the form of a multivariate polynomial of the energies without
any fraction.
\\The previous ideas help us suggest other computational algorithm strategies. We call this method algorithmic of finite and high-temperature series (AFHTS).
\section{The algorithm procedure of AFHTS}
In summary, the algorithmic steps of the calculation of free energy at finite and high-temperature are illustrated as follows:
\begin{enumerate}
	\item Generate all essentially distinct Hugenholtz vacuum type diagrams of order $n$, in this case we use our previous fast algorithm Ref.~[\onlinecite{tag2017}]. Here the total number of vertices $N_V$ and edges $N_E$ of each generated diagram $G(V,E)$ is $N_V=n$ and $N_E=2n$ respectively. The edges of diagram $G(V,E)$ are labeled as unsigned integers;
	\item Use depth-first search (DFS) to find an arbitrary spanning tree ${E_{s_1},E_{s_2},...,E_{s_{n-1}}}$ of the diagram $G$, we call this tree the divisor tree $DT$. The set of $DT$ are arranged in ascending order $s_1<s_2<...<s_{n-1}$;
	\item Relabel the vertices of diagram $G$ according to order of encountered vertices in DFS search, this step help us to ensure that each vertex $v_i$ of the diagram $G$ is connected with the previous vertex $v_{i-1}$;
	\item Represent each edge of the diagram $G(E,V)$ as a binary forms, i.e., $E_k=2^{k-1}$;
	\item Select the first branch $s_1$ from $DT$ and generate all spanning trees whose branches do not contain $s_1$, but with the following condition: If the generated spanning tree contain some branches of $DT$, then the fundamental cut of those branches must cross one or more of the remain chords of $DT$;
	\item Using the partial fractions in which the sets of denominators $D_1,...,D_{k_1}$ contains the branch $s_1$ and $D_{k_1+1},...,D_{k_2}$ contains the branch $s_2$,…etc.
\end{enumerate}
The contribution of each generated spanning tree from $G$ can be formulated as:
\begin{equation}
    \Omega_{n}^{G}=f^{[O_1]}(E_1)...f^{[O_s]}(E_s)\prod_{k=1}^{M}f^{-1}[E_{s_k},d_1,...,d_{r_k}]
    \label{eq:25}.
\end{equation}
The integer number $r_k=n_k-m_k$ is the rank of energy branch $E_{s_k}\in DT$, where $n_k$ is the number of times the energy $E_{s_k}$ appears in the denominator $D=D_1D_2...D_{n-1}$ whereas $m_k$ is the number of presence of $E_{s_k}$ with one of the previous branches ${E_{s_1},E_{s_2},...,E_{s_{k-1}}}$ in the same denominator $D_i$. All ranks $r_k$ fulfills the following condition
\begin{equation}
    \sum_{k=1}^{n-1}r_k=n-1
    \label{eq:26}.
\end{equation}
The amount of $ff$ is the divided difference \cite{deBoor2005}, which is a mathematical tool used certainly to calculate the coefficients in the interpolation polynomial in the Newton form. The recursive relation of divided difference is defined by:
\begin{equation}
    f^{-}[e_s]=f^{[O_i]}(e_s)
    \label{eq:27}.
\end{equation}
\begin{eqnarray}
\begin{aligned}
f^-[e_s,d_1]&=\frac{f^{-}(e_s)-f^{-}(d_1)}{e_s-d_1}
\\
&=\frac{1}{2}\left(\frac{\tanh(\frac{\beta}{2} e_s)-\tanh(\frac{\beta}{2}d_1)}{e_s-d_1} \right)
\end{aligned}
\label{eq:28}.
\end{eqnarray}
And for $r\geq 2$
\begin{equation}
    f^-[e_s,...,d_r]=\frac{f^-[e_s,...,d_{r-1}]-f^-[d_1,...,d_r]}{e_s-d_r}
    \label{eq:29}.
\end{equation}
Now we can find the high-temperature series expansion of the divided difference Eq.~(\ref{eq:29}) in the form of an infinite sum of the order $\beta$ directly 
\begin{equation}
    f^-[e_s,...,d_r]=-\frac{1}{2}\sum_{n=r}^{\infty}\frac{\beta^n}{n!}e_n\sum_{k_0k_1...k_r}e_{s}^{k_0}d_{1}^{k_1}...d_{r}^{k_r}
    \label{eq:30}.
\end{equation}
Here $d_j$ is the denominator $D_j$ without $e_s$, i.e $d_j=e_s\mp D_j$, where the sign – (+) indicates if $e_s$ is in the same direction (opposite direction) of the branch $e_s$. 
\\The coefficients $e_n$ are the Euler numbers defined by the relation Eq.~(\ref{eq:19}).
\\Finally, the integers $k_i\geq 0$ are selected according to the following relation:
\begin{equation}
    \sum_{i=0}^{r}k_i=n-r
    \label{eq:31}.
\end{equation}
As we see the direct application of relation Eq.~(\ref{eq:30}) can be evaluate the contribution Eq.~(\ref{eq:25}) in high temperature very efficiently. 
\\In finite temperature, the relation Eq.~(\ref{eq:29}) can be used to calculate the contribution Eq.~(\ref{eq:25}) by numerically integrating and interpolating the divided difference Eq.~(\ref{eq:29}) in recursive manner.
\\By applying the suggested algorithm, we can formulate any desired order of $\Omega$ as a divided difference. For this purpose, we provide the series of $\Omega$ up to order 6 in the supplemental material.
\\One of the biggest benefit of this method is when we can formulate some types of diagrams in one recursive relation. For this, we found that the contribution of Hugenholtz ladder type diagrams $\Omega^{(L)}$ (Like graphs $G_{2}^{1}$, $G_{3}^{1}$ and $G_{4}^{3}$ in Fig.~(\ref{fig:img0})) and the contribution Hugenholtz loop type diagrams (Like graphs $G_{3}^{2}$ and $G_{4}^{4}$ in Fig.~(\ref{fig:img0})) can be formulated as:
\begin{equation}
\Omega^{LO}=\sum_{n=2}^{\infty}\frac{1}{n}\int_{P}V_{n}^{+}\Omega_{n}^{+} dP_n+\sum_{n=3}^{\infty}\frac{1}{n}\int_P V_{n}^{-}\Omega_{n}^{-}dP_n
\label{eq:L32}
\end{equation}
Where the general formula of ladder $V_{n}^{+}$ and loop $V_{n}^{-}$ potentials defined as
\begin{eqnarray}
\begin{aligned}
V_{n}^{+}&=\frac{1}{4^n}V_{p_{1},p_{2}}^{p_{2n-1},p_{2n}}\prod_{i=1}^{n-1}V_{p_{2i+1},p_{2i+2}}^{p_{2i-1},p_{2i}}
\\
V_{n}^{-}&=\frac{1}{2^{n+1}}V_{p_{1},p_{2n}}^{p_{2n-1},p_{2}}\prod_{i=1}^{n-1}V_{p_{2i+1},p_{2i}}^{p_{2i-1},p_{2i+2}}
\label{eq:L33}
\end{aligned}
\end{eqnarray}
The contribution of ladder $\Omega_{n}^{+}$ and loop $\Omega_{n}^{-}$ in the grand potential generalized as follows
\begin{equation}
\Omega_{n}^{\pm}=\prod_{i=1}^{n}\left(\tanh\left(\frac{\beta}{2}E_{p_{2i-1}}\right)\pm \tanh\left(\frac{\beta}{2}E_{p_{2i}}\right)\right)B[x_{1}^{\pm},x_{2}^{\pm},...,x_{n}^{\pm}]
\label{eq:L34}
\end{equation}
Where the variable $x_{i}^{\pm}, 1\leq i\leq n$ defined as
\begin{equation}
x_{i}^{\pm}=E_{p_{2i}}\pm E_{p_{2i-1}}
\label{eq:L35}
\end{equation}
The general function of a divided difference $B(x)$ is defined as a Bose-Einstein distribution:
\begin{equation}
B(x)=\frac{1}{e^{\beta x}-1}
\label{eq:L36}
\end{equation}
The integration in Eq.~(\ref{eq:L32}) is over all momentum variables $dP_n=dp_1dp_2...dp_{2n}$

\section{High-temperature series expansion of Heisenberg XXZ Spin 1/2 model}
\begin{figure*}[h]
   \centering
\begin{tabular}{lccccc}
\includegraphics[scale = 0.8]{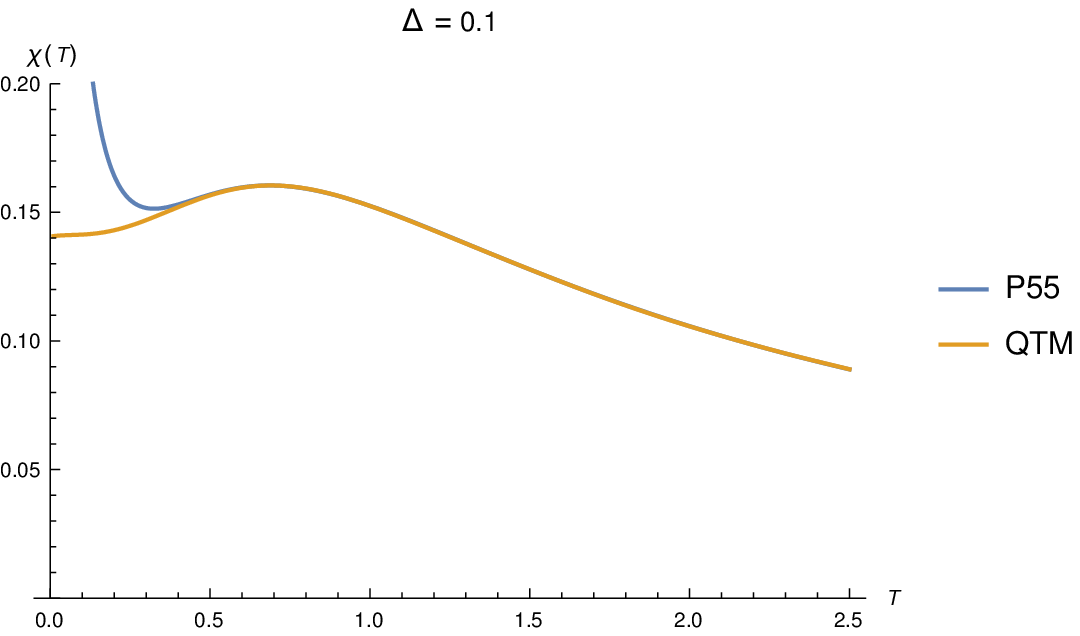}&
\includegraphics[scale = 0.8]{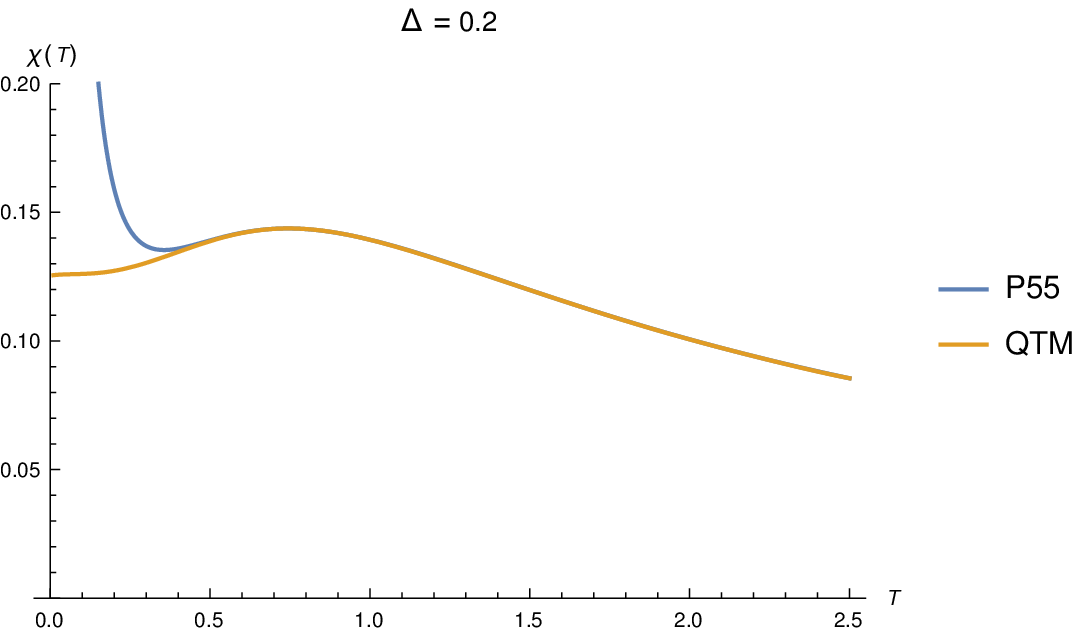}\\
\\
\includegraphics[scale = 0.8]{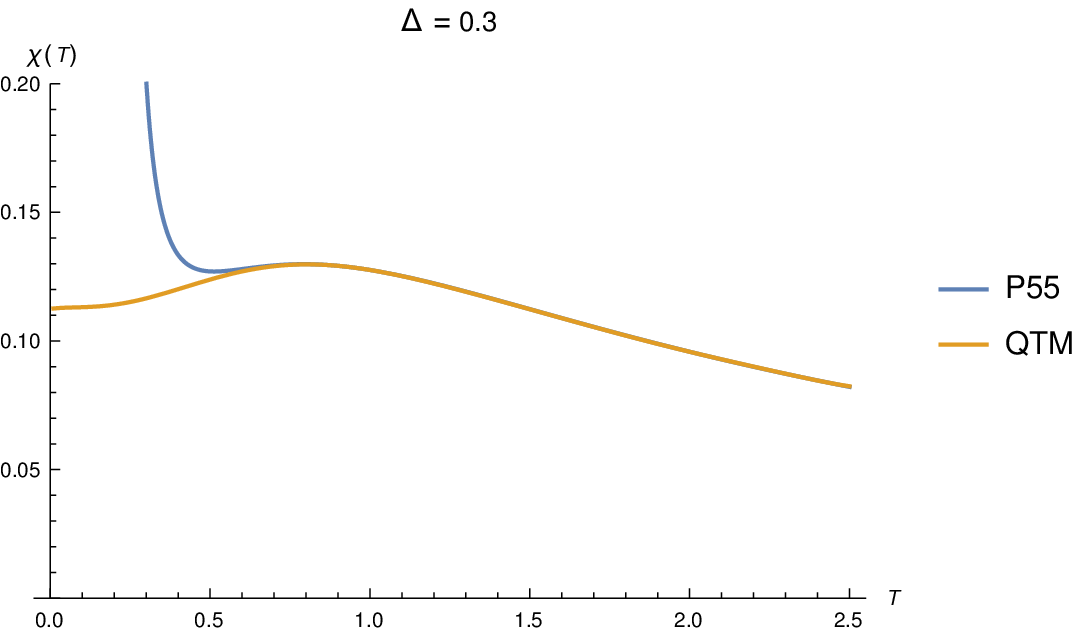}&
\includegraphics[scale = 0.8]{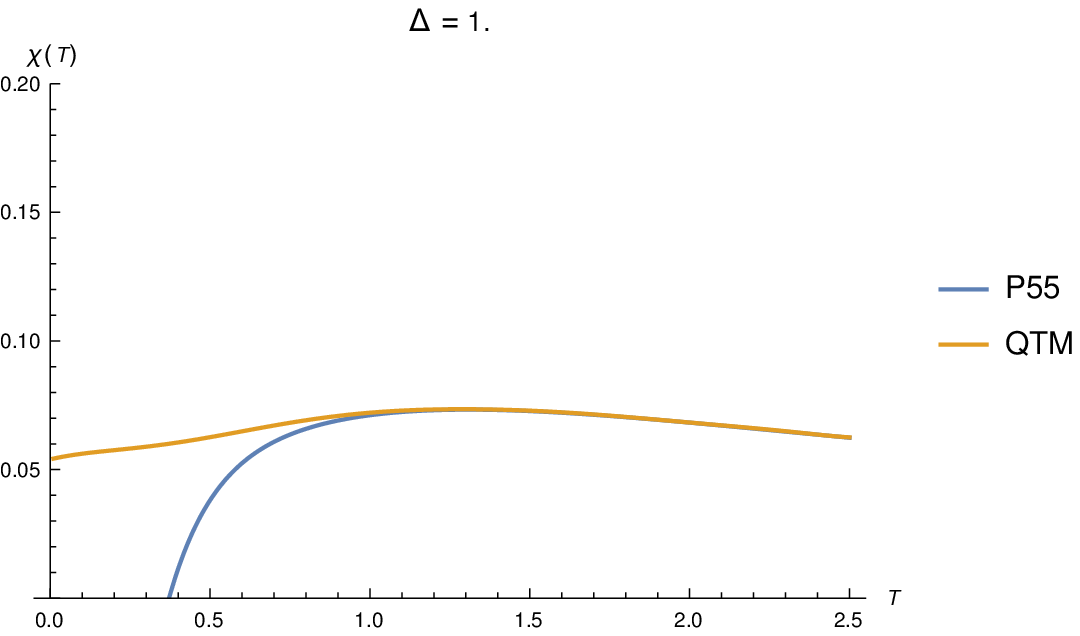}
\end{tabular}
    \caption{Magnetic susceptibility of the Heisenberg XXZ chain for $\Delta=0.1$, $\Delta=0.2$, $\Delta=0.3$ and $\Delta=1.0$ at $h=0$}
    \label{fig:figu2} 
\end{figure*}

The simplest Hamiltonian of the interacting model is the Heisenberg spin $1/2$ chain. In the general case, the one-dimensional XXZ model in the presence of external magnetic field $h$ can be written as follows:
\begin{equation}
    H=-\frac{J}{2}\sum_{k=1}^{N}(S_{k}^{+}S_{k+1}^{-}+S_{k}^{-}S_{k+1}^{+})-J\Delta \sum_{k=1}^{N}S_{k}^{z}S_{k+1}^{z}-2h\sum_{k=1}^{N}S_{k}^{z}
    \label{eq:32}.
\end{equation}
Where $S_k^+=S_k^x+i S_k^y$, $S_k^+=S_k^x-i S_k^y$. Using the Jordan-Wigner transform Ref.~[\onlinecite{Jorden1928}], the Hamiltonian Eq.~(\ref{eq:32}) can be written in terms of the fermionic operators $ c_{k} $ and $ c_{k}^+ $. By applying the Fourier transform to these operators $c_k=\frac{1}{\sqrt{N}}\sum_{p}e^{i\frac{2\pi}{N}pka}c_p$, $c_{k}^+=\frac{1}{\sqrt{N}}\sum_{p}e^{-i\frac{2\pi}{N}pka}c_{p}^+$, the Hamiltonian Eq.~(\ref{eq:28}) can therefore expressed as a same form of the two-body interacting systems Eq.~(\ref{eq:1}). Where, the quasiparticles energy $\varepsilon_p$ and the potential $V_{m,l}^{r,s}$ written as:
\begin{equation}
    \varepsilon_{p}=J\Delta-2h-J \cos(p)
    \label{eq:33}.
\end{equation}
\begin{equation}
V_{m,l}^{r,s}=-4J\Delta \sin \left(\frac{s-r}{2}\right)\sin \left(\frac{l-m}{2}\right)\delta_{r+s,m+l}        
\label{eq:34}.
\end{equation}
Where $\delta_{r+s,m+l}$ represents the Kronecker delta and $p,r,s,m,l\in [-\pi,\pi]$.
\\By applying the potential Eq.~(\ref{eq:34}) in Hartree-Fock equation Eq.~(\ref{eq:6}), the energy Eq.~(\ref{eq:5}) can be formulated as
\begin{equation}
    E_p(\beta)=A(\beta)+B(\beta)\cos(p)
    \label{eq:35}.
\end{equation}
Where:
\begin{equation}
  A(\beta)=-2h+\frac{J\Delta}{2\pi}\int_{-\pi}^{\pi}\tanh\left(\frac{\beta}{2}E_q(\beta)\right)dq  
  \label{eq:36}.
\end{equation}
\begin{equation}
  B(\beta)=-J-\frac{J\Delta}{2\pi}\int_{-\pi}^{\pi}\tanh\left(\frac{\beta}{2}E_q(\beta)\right)\cos(q)dq 
  \label{eq:37}.
\end{equation}
In high-temperature expansion, we assume $A(\beta)=\sum_{i=0}^{\infty}a_{i}\beta^{i}$ and $B(\beta)=\sum_{i=0}^{\infty}b_{i}\beta^{i}$ and substitute them into the equations Eq.~(\ref{eq:36}) and Eq.~(\ref{eq:37}). Then we can get the equations that determine the series $A(\beta)$ and $B(\beta)$ recursively by comparing the same order of $\beta$ in LHS and RHS. For example, we have $a_0=-2h$, $b_0=-J$, $a_1=-J h \Delta$, $b_1=\frac{\Delta J^2}{4}$,...In this manner, we can calculate any order of Hartree-Fock energy Eq.~(\ref{eq:35}) with a simple code in Mathematica.
\\The high-temperature expansion of the free energy can be calculated directly using the algorithm described in section IV. Each contribution $\Omega_{n}^{G_i}$ of Hugenholtz diagrams $G$ at order $n$ is a multivariate polynomial of Hartree-Fock energies $E_{p_1},E_{p_2},...,E_{p_{2n}}$ and potential $\prod_{i=1}^{n}V_{m_i,l_i}^{r_i,s_i}$. In fact, the remaining variables in $\Omega_{n}^{G_i}$ are $n + 1$ among $2n$. This reduction results from applying the Kronecker delta  $\delta_{r_i+s_i,m_i+l_i}$, but the traditional integrating methods have difficulty when using it to integrate overall $n+1$ variables in the domains $[-\pi,\pi]$. In this case, we use the following method to simplify the integration.
\\We can write the expression of Hartree-Fock energy Eq.~(\ref{eq:35}) and potential Eq.~(\ref{eq:34}) in the following forms:
\begin{equation}
    E_{p}=A(\beta)+\frac{B(\beta)}{2}(z_p+z_{p}^{-1})
    \label{eq:38}.
\end{equation}
\begin{equation}
V_{m,l}^{r,s}=J\Delta (z_r-z_s)(z_m-z_l)(z_{r}z_{s}z_{m}z_{l})^{-1/2}\delta_{r+s,m+l}
\label{eq:39}.
\end{equation}
Where $z_k=e^{i k}$. After replacing the relation of Eq.~(\ref{eq:38}) and Eq.~(\ref{eq:39}) in the expression of $\Omega_{n}^{G_i}$, the latter will be a multivariate polynomial in terms of variables $z_{p_1},z_{p_1},...,z_{n+1}$. When integrating $\Omega_{n}^{G_i}$ on a variable, say ${p_1}$, it suffices to find the constant coefficient $a_0$ of the polynomial  $P(z_{p_1})=\sum_{i=0}a_i z_{p_1}^i$. Of course this coefficient contains the other variables. In other words:
\begin{equation}
    \frac{1}{2\pi}\int_{-\pi}^{\pi}\Omega_{n}^{G_i}(z_{p_1},z_{p_2},...,z_{p_{n+1}})d p_{1}=a_0(z_{p_2},...,z_{p_{n+1}})
    \label{eq:40}.
\end{equation}
By the same way, we repeat the integration procedure Eq.~(\ref{eq:40}) on the other variables $p_2,...,p_{n+1}$. This method is faster than ordinary integration because we do not need to expand the expression of $\Omega_{n}^{G_i}$, which takes a lot of time and space.
\\Finally, we have found the sixth order of free energy at high temperature by implementing the above algorithm and the integration procedure Eq.~(\ref{eq:40}) in Mathematica.
\begin{widetext}
\begin{eqnarray}
&\Omega=-\frac{\log (2)}{\beta }+\frac{J \Delta}{4}+\beta  \left(-\frac{h^2}{2}-\frac{J^2}{32} \left(\Delta ^2+2\right) \right)+\frac{J \Delta \beta ^2 }{2!} \left(\frac{J^2}{32}-\frac{h^2}{2}\right)+\frac{\beta ^3}{3!}\left(\frac{h^4}{2}-\frac{3}{8} \left(\Delta ^2-1\right) h^2 J^2+\frac{J^4}{512}\left(\Delta ^4+8 \Delta ^2+6\right) \right)+\nonumber\\
&\frac{J \Delta \beta ^4 }{4!}\left(4 h^4-\frac{h^2 J^2}{4} \left(\Delta ^2-3 \right) -\frac{3 J^4}{128} \left(\Delta ^2+2 \right) \right)+\nonumber\\
&\frac{\beta ^5}{3 ( 5!)}\left(-8 h^6+\frac{15 h^4 J^2}{4}\left(13 \Delta ^2-4 \right)-\frac{15 h^2 J^4}{32}\left(\Delta ^4-2 \Delta ^2+6 \right)-\frac{J^6}{1024}\left(2 \Delta ^6+36 \Delta ^4-15 \Delta ^2+40 \right) \right)+\nonumber\\
&\frac{J \Delta \beta ^6 }{6!}\left(-68 h^6+\frac{5 h^4 J^2}{4} \left(40 \Delta ^2-39 \right)-\frac{3 h^2 J^4}{32}\left(\Delta ^4+5 \Delta ^2+15 \right) +\frac{J^6}{1024} \left(34 \Delta ^4+194 \Delta ^2+165 \right) \right)
\label{eq:41}.
\end{eqnarray}
\end{widetext}
The result of Eq.~(\ref{eq:41}) for $\Delta = 1$ (XXX Heisenberg chain) entirely coincides with the works of Ref.~[\onlinecite{takahashi2002}]. We also remark that the coefficients up to order $\beta ^3$ agree with those given in Ref.~[\onlinecite{Rojas2002}].
\\We show in Fig.~\ref{fig:figu2} the plain series (PS) of magnetic susceptibility $\chi=-\frac{\partial ^2 \Omega }{\partial h^2}$ at zero magnetic field up to order 10 with the application of standard Padé approximation [5,5] (PD55) for our PS. We compare the numerical solution of $\chi$ obtained from the exact method QTM (Quantum Transfer Matrix) Ref.~[\onlinecite{Klumber2004}]. From Fig.~\ref{fig:figu2}, we find that the Padé approximation shows a good coincidence with data of QTM. We also note that the relative error between QTM and PD55 increases with increasing $\Delta$ values, but PD55 follows the same curve as QTM.
\section{Conclusion}
We have presented a general algorithm to calculate Matsubara sums for any Hugenholtz or Feynman vacuum diagrams and formulate the results as a divided difference. The importance of our method is that in addition to applying in any interaction, it allows us to calculate thermodynamic quantities at high and finite temperatures in any dimension.
\\Our method enables us to dispense with the traditional integration methods, such as the Monte Carlo method, which requires a significant computational time. We need only combine our HTS Eq.~(\ref{eq:26}) and Eq.~(\ref{eq:30}) with the Padè approximation to find very accurate numerical results for physical quantities at sufficiently low temperatures. Additionally, we can apply the direct formula of divided difference Eq.~(\ref{eq:29}) to find all needed results at $T=0$.
\\As proof of concept, we presented the application of this method to the spin-1/2 Heisenberg XXZ chain model and derived its free energy per site $\Omega$ up to sixth order. The resulting series of $\Omega$ agree with the work of Ref.~[\onlinecite{Rojas2002}] and for d=1 with Ref.~[\onlinecite{takahashi2002}]. We also compared our data with other numerical methods  Ref.~[\onlinecite{Klumber2004}] we found excellent results.
\\Of course, our method applies to the evaluation of self-energy, and we can formulate it as a divided difference; we will study such possibilities in the forthcoming work.
\nocite{*}

\bibliography{references}

\end{document}